\documentclass[12pt]{article}
\usepackage{epsfig}
\usepackage{amsfonts}
\usepackage{amscd}
\usepackage{latexsym}
\usepackage{amsmath,amssymb}
\usepackage{verbatim}
\usepackage{setspace}
\usepackage{color}
\usepackage{cite}

\usepackage[textheight=9in, textwidth=6.5in, letterpaper]{geometry}

\usepackage{hyperref}
\hypersetup{
    colorlinks,
    citecolor=black,
    filecolor=black,
    linkcolor=black,
    urlcolor=black
    linktoc=all
}

\numberwithin{equation}{section}

\def\ip{{\mathcal I}^+}
\def\im{{\mathcal I}^-}

\def\e{{\epsilon}}

 \def\p{\partial}

\def\0{{(0)}}
\def\1{{(1)}}
\def\2{{(2)}}
\def\a{{\alpha}}

\def\cO{{\cal O}}

\def\ci{{\mathcal I}}

\def\<{\langle }
\def\>{\rangle }

\newcommand{\bea}{\begin{eqnarray}}
\newcommand{\eea}{\end{eqnarray}}
\newcommand{\be}{\begin{equation}}
\newcommand{\ee}{\end{equation}}
\newcommand{\ba}{\begin{align}}
\newcommand{\ea}{\end{align}}
\def\be{\begin{equation}}
\def\ee{\end{equation}}
\def\beq{\be\begin{array}{c}}
\def\eeq{\end{array}\ee}
\def\Phi_1{E_r }

\newcommand{\bigO}{\mathcal{O}}

\newcommand{\w}{z}

\renewcommand{\epsilon}{\varepsilon}

   \makeatletter
  \let\over=\@@over \let\overwithdelims=\@@overwithdelims
  \let\atop=\@@atop \let\atopwithdelims=\@@atopwithdelims
  \let\above=\@@above \let\abovewithdelims=\@@abovewithdelims
\renewcommand\section{\@startsection {section}{1}{\z@}%
                                   {-3.5ex \@plus -1ex \@minus -.2ex}
                                   {2.3ex \@plus.2ex}%
                                   {\normalfont\large\bfseries}}

\renewcommand\subsection{\@startsection{subsection}{2}{\z@}%
                                     {-3.25ex\@plus -1ex \@minus -.2ex}%
                                     {1.5ex \@plus .2ex}%
                                     {\normalfont\bfseries}}

\begin{document}
\begin{titlepage}
\unitlength = 1mm
{\begin{flushright} CALT-TH-2015-006 \end{flushright}}
\ \\
\vskip 1cm
\begin{center}

{ \LARGE {\textsc{Higher-Dimensional Supertranslations and Weinberg's Soft Graviton Theorem}}}

\vspace{0.8cm}
Daniel  Kapec$^\dagger$, Vyacheslav Lysov$^*$, Sabrina Pasterski$^\dagger$ and Andrew Strominger$^\dagger$

\vspace{1cm}

$^\dagger${\it  Center for the Fundamental Laws of Nature, Harvard University,\\
Cambridge, MA 02138, USA}\\

$^*${\it Walter Burke Institute for Theoretical Physics,\\
 California Institute of Technology, \\
 Pasadena, CA 91125, USA}

\begin{abstract}
Asymptotic symmetries of theories with gravity in $d=2m+2$ spacetime dimensions are reconsidered for $m>1$ in light of recent results concerning  $d=4$ BMS symmetries. Weinberg's soft graviton theorem in $2m+2$ dimensions is re-expressed as a Ward identity for the gravitational $\mathcal{S}$-matrix. The corresponding asymptotic symmetries are identified with $2m+2$-dimensional supertranslations.  An alternate derivation of these asymptotic symmetries as diffeomorphisms which preserve finite-energy boundary conditions at null infinity and act non-trivially on physical data is given.   
Our results  differ from those of previous analyses whose stronger boundary conditions precluded  supertranslations for $d>4$.  We find for all even $d$ that supertranslation symmetry is spontaneously broken in the conventional vacuum and identify soft gravitons as the corresponding Goldstone bosons.

\end{abstract}

\vspace{1.0cm}

\end{center}

\end{titlepage}

\pagestyle{empty}
\pagestyle{plain}

\def\vx{{\vec x}}
\def\p{\partial}
\def\po{$\cal P_O$}

\pagenumbering{arabic}

\tableofcontents
\section{Introduction}
Asymptotic symmetry groups play a vital role in our modern understanding of general relativity. Although the concept originated in the early 1960's, it continues to influence much of the contemporary research on classical and quantum gravity. From holography and black hole thermodynamics to the infrared behavior of the gravitational $\mathcal{S}$-matrix, asymptotic symmetry groups have provided crucial  insights into many of today's most exciting research topics. They will undoubtedly continue to play a critical role in further clarifying the nature of quantum gravity. 

The asymptotic symmetry group of asymptotically flat spacetimes is particularly interesting from both theoretical and phenomenological points of view. Early research on this topic by Bondi, Van der Berg, Metzner and Sachs \cite{Bondi:1962px,Sachs:1962wk,Sachs:1962zza} led to a surprising conclusion: the asymptotic symmetry group of four dimensional asymptotically flat spacetimes is not the finite-dimensional Poincare group, but an infinite dimensional group now known as the Bondi-Metzner-Sachs (BMS) group.
The BMS group contains the boosts, rotations and translations that comprise the isometry group of flat spacetime. However, it also includes an infinite dimensional abelian subgroup, known as the supertranslation subgroup, whose existence seems to have troubled the early pioneers of the subject. Repeated attempts to eliminate these extra symmetries proved unsuccessful, and the BMS group ultimately gained acceptance as the physically correct asymptotic symmetry group for four dimensional asymptotically flat spacetimes. 

It was eventually recognized that the BMS supertranslations were related to the infrared behavior of the gravitational theory \cite{Ashtekar:1987tt,Ashtekar:1981sf,Ashtekar:2014zsa}.  This relationship has recently been made precise. In \cite{Strominger:2013jfa} it was argued that a certain diagonal subgroup of the product of the past and future BMS groups is a symmetry of both classical and quantum  gravitational scattering. In  \cite{He:2014laa} it was further demonstrated that the Ward identity associated to this diagonal supertranslation invariance of the gravitational $\mathcal{S}$-matrix is equivalent to Weinberg's soft graviton theorem \cite{Weinberg:1965nx}. Further investigations along these lines have established a robust and detailed correspondence between soft theorems for gauge theory/gravity scattering amplitudes and Ward identities for extended asymptotic symmetry groups \cite{Strominger:2013jfa, He:2014laa, Strominger:2013lka,  Kapec:2014opa,He:2014cra, Lysov:2014csa, Kapec:2014zla,Mohd:2014oja, Larkoski:2014bxa,Campiglia:2014yka,Campiglia:2015yka, Banks:2014iha,Adamo:2014yya,Geyer:2014lca,Bern:2014vva}. Moreover, it has been shown \cite{Strominger:2014pwa} that the gravitational memory effect \cite{ZeldPoln, bragthorne,ChristodoulouCR}, which occurs in the deep infrared,  provides direct and measurable consequences of  BMS symmetry. 

Although the asymptotic symmetry group of asymptotically flat spacetimes is well-studied in four dimensions, the higher dimensional analog has received limited attention \cite{Hollands:2003xp,Hollands:2003ie,Tanabe:2009va,Tanabe:2010rm,Tanabe:2011es,Tanabe:2012fg,Barnich:2006av,Awada:1985by}.  Interestingly, nearly all of these analyses concluded that supertranslations do not exist in higher dimensions. This result seems to be at odds with the soft theorem/asymptotic symmetry correspondence, given that Weinberg's soft graviton theorem holds in any dimension. The resolution of this discrepancy is the focus of this paper. 

Briefly, the analyses that claim to eliminate supertranslations in higher dimensions do so by placing restrictive boundary conditions on the metric at null infinity. We demonstrate that by slightly relaxing these boundary conditions to allow for zero-energy, ``large diffeomorphism"  contributions to the metric, one may recover the full BMS group, including supertranslations,  in any even dimension.\footnote{In $d=4$, the `boundary gravitons' produced by supertranslations and the radiative gravitons both appear at the same order in ${1 \over r}$. For higher $d$, the boundary gravitons appear at lower order in the $1 \over r$ expansion, so that 
boundary conditions constraining pure diffeomorphisms at lower order than the radiative modes will kill supertranslations. 
The situation here is like $AdS_3$ \cite{Brown:1986nw}, where the leading allowed excitations are all pure diffeomorphisms.} 
We corroborate this with an investigation of Weinberg's soft graviton theorem in arbitrary even-dimensional spacetimes. Manipulation of this universal relation allows us to prove a Ward identity relating $\mathcal{S}$-matrix elements and derive a corresponding conserved charge. This charge 
is rewritten, using the constraints and our boundary conditions,  as a boundary integral involving the generalized Bondi mass aspect.  A set of brackets is proposed for which this charge generates the supertranslations. Hence, our weakened boundary conditions allow a derivation of the Weinberg soft identities between $\mathcal{S}$-matrix elements, while the imposition of stronger boundary conditions miss this important feature of scattering. 
The argument can also be turned around, reverse-engineering  Weinberg's soft theorem into a supertranslation symmetry of gravitational scattering. 

We defer the study of odd dimensions due to known difficulties in defining null infinity in odd-dimensional spacetimes \cite{Hollands:2004ac}. 

The outline of this paper is as follows:  in section 2 we define and discuss asymptotically flat spacetimes in even dimensions. We establish our coordinate system and relevant boundary conditions, and then derive the corresponding asymptotic symmetry group. In section 3 we briefly discuss the semiclassical gravitational scattering problem as well as the known subtleties in connecting past and future null infinity.  In section 4 we specialize to six dimensions for ease of presentation, deriving equations needed in the analysis of Weinberg's soft theorem. In section 5 we derive a Ward identity from Weinberg's soft theorem, and in section 6 we demonstrate its equivalence to the supertranslation  Ward identity for soft gravitons and hard matter fields. We do $not$ verify that the terms in the charge which are quadratic in the metric perturbations correctly generate supertranslations for hard gravitons. We expect this to be the case but an analysis of gauge fixing and Dirac brackets at subleading order would be required. Section 7 details the generalization to arbitrary even-dimensional spacetime. 

\section{General Relativity in $d=2m+2$ dimensions   }

In this section we study even-dimensional asymptotically flat solutions to Einstein's field equations without a cosmological constant. Working in Bondi gauge, we propose appropriate boundary conditions for asymptotically flat spacetimes and derive their asymptotic symmetry groups. Our definition of asymptotic flatness differs from that used in previous analyses, and we comment on the implications. Finally, we collect a series of useful equations in the linearized theory.

\subsection{Bondi gauge}
Metric solutions to Einstein's equations in $d=2m+2$ dimensions satisfy
\be
R_{\mu\nu}-\frac12 R g_{\mu\nu} =8\pi G T_{\mu\nu}^M,   
\ee
where $T_{\mu\nu}^M$ is the matter stress-energy tensor. We choose coordinates $u,r,z^a$ ($a=1,\dots,2m$) that asymptote to the usual retarded coordinates on flat spacetime. At large - $r$ and in terms of flat space Cartesian coordinates $t,x^i$ we have 
\be \label{embed}
u=t-r, \;\;\;\;\; r^2=x^ix_i, \;\;\;\;\; x^i=r\hat{x}^i(z^a),   
\ee
where $\hat{x}^i(z^a)$ defines an embedding of $S^{2m}$ in $\mathbb{R}^{2m+1}$. Future null infinity $\mathcal{I}^+$ is given by the null hypersurface $(r=\infty,u,z^a)$,  with future ($u=\infty$) and past ($u=-\infty$) boundaries denoted by $\mathcal{I}^+_+$ and $\mathcal{I}^+_-$, respectively. In this coordinate system,  Bondi gauge is defined by the $2m+2$ gauge-fixing conditions  
\be \label{bondigaugemet}
g_{rr}=0, \hspace{.5 in} g_{ra}=0, \hspace{.5 in}\det g_{ab}=r^{4m}\det{\gamma_{ab}},   
\ee
 where $\gamma_{ab}$ is the round metric on the unit $S^{2m}$ with covariant derivative $D_a$. Such a metric can always be put into the form  
\be
\label{eq:BM}
ds^2=e^{2\beta}M du^2-2e^{2\beta}dudr+g_{ab}(dz^a-U^adu)(dz^b-U^bdu).   
\ee
For the case of asymptotically flat spacetimes, we will assume $\beta$, $M$, $U_a$, and $g_{ab}$ admit an expansion near $\mathcal{I}^+$ of the form:
\begin{gather} \label{eq:e1} \notag
\beta=\sum_{n=2}^{\infty}\frac{\beta^{(n)}(u,z)}{r^n},  \hspace{.5 in} M=-1+\sum_{n=1}^{\infty}\frac{M^{(n)}(u,z)}{r^n},  \hspace{.5 in} U_a=\sum\limits_{n=0}^\infty \frac{U_a^{(n)}(u,z)}{r^n},
\\  g_{ab}=r^2\gamma_{ab}+\sum_{n=-1}^{\infty}\frac{C_{ab}^{(n)}(u,z)}{r^n}.   
\end{gather}
In the vicinity of past null infinity, $\mathcal{I}^-$, we choose advanced coordinates $v,r,z^a$ asymptotically related to the flat space Cartesian coordinates through the relations
\be
v=t+r, \;\;\;\;\; r^2=x^ix_i, \;\;\;\;\; x^i=-r\hat{x}^i(z^a).   
\ee
Here $\hat{x}^i$ is the same embedding of the $S^{2m}$ in $\mathbb{R}^{2m+1}$ as in (\ref{embed}). Note in particular that the angular coordinate $z^a$ on $\mathcal{I}^-$ is antipodally related to the angular coordinate on $\mathcal{I}^+$, so that null generators of $\mathcal{I}$ passing through spatial infinity ($i^0$)  are labeled by the same numerical value of $z^a$. $\mathcal{I}^-$ is the $(r=\infty,v,z^a)$ null hypersurface, with future $(v=\infty)$ and past $(v=-\infty)$ boundaries denoted by $\mathcal{I}^-_+$ and $\mathcal{I}^-_-$, respectively. The metric in advanced Bondi gauge takes the form
\be
\label{eq:BM}
ds^2=e^{2\beta^-}M^- dv^2+2e^{2\beta^-}dvdr+g^-_{ab}(dz^a-W^adv)(dz^b-W^bdv),   
\ee
where $\beta^-$, $M^-$, $W_a$, and $g^-_{ab}$ admit the large-$r$ expansions 
\begin{gather}\label{eq:e2} \notag
\beta^-=\sum_{n=2}^{\infty}\frac{\beta^{-(n)}(v,z)}{r^n}, \hspace{.5 in}M^-=-1+\sum_{n=1}^{\infty}\frac{M^{-(n)}(v,z)}{r^n},  \hspace{.5 in} W_a=\sum\limits_{n=0}^\infty \frac{W_a^{(n)}(v,z)}{r^n},  
\\
g^-_{ab}=r^2\gamma_{ab}+\sum_{n=-1}^{\infty}\frac{D_{ab}^{(n)}(v,z)}{r^n}.   
\end{gather}

\subsection{Asymptotically flat spacetimes}
Having fixed a coordinate system, we must now choose the boundary conditions that define asymptotic flatness at $\mathcal{I}^+$ in this gauge. Our conditions on the metric components are the same as those in four dimensions: 
\be \label{bc1}
g_{uu}=-1+\bigO(r^{-1}), \hspace{.3 in} g_{ur}=-1+\bigO(r^{-2}), \hspace{.3 in} g_{ua}=\bigO(1), \hspace{.3 in} g_{ab}=r^2\gamma_{ab}+\bigO(r).   
\ee
We also require
\be \label{bcric} 
R_{uu}=\bigO(r^{-2m}), \hspace{.5 in} R_{ur}=\bigO(r^{-2m-1}), \hspace{.5 in} R_{ua}=\bigO(r^{-2m}),  
\ee
\be \label{bc2}
R_{rr}=\bigO(r^{-2m-2}), \hspace{.5 in} R_{ra}=\bigO(r^{-2m-1}), \hspace{.5 in}   R_{ab} = \bigO(r^{-2m}).   
\ee
The analogous conditions define asymptotic flatness at $\mathcal{I}^-$. When the theory is coupled to matter sources, we impose the same falloff conditions on the components of $T_{\mu \nu}^M $ as on $R_{\mu \nu}$. It is important to note that the boundary conditions (\ref{bc1})-(\ref{bc2}) are less restrictive than those typically considered in the literature \cite{Hollands:2003xp,Hollands:2003ie,Tanabe:2009va,Tanabe:2010rm,Tanabe:2011es,Tanabe:2012fg}. In particular, the falloff condition on $g_{ab}$ in equation (\ref{bc1}) does not reflect the large-$r$ behavior of generic gravitational radiation in $d=2m+2$ dimensions, for which $g_{ab}=r^2\gamma_{ab}+\bigO(r^{2-m})$. As we will see, the choice of this boundary condition essentially determines whether or not the corresponding asymptotic symmetry group contains supertranslations.  Naively, our less restrictive falloff conditions on the metric components could lead to bad behavior at infinity and divergences in physical quantities. However, the metric itself is not physically observable and the boundary conditions on the Ricci tensor ensure finiteness of energy flux and other gravitational observables. As we will see, the potentially dangerous pieces of the metric turn out to be pure ``large diffeomorphism" for the spacetimes we consider. In the next section we demonstrate that allowing these leading pieces of the metric to be pure diffeomorphism, rather than setting them equal to zero, enlarges the asymptotic symmetry group from $ISO(2m+1,1)$ to $BMS_{2m+2}$.

\subsection{Asymptotic symmetries}
We are now in a position to discuss the asymptotic symmetry group of asymptotically flat $2m+2$ dimensional spacetimes. We define the asymptotic symmetry group to be the group of all diffeomorphisms preserving Bondi gauge (\ref{bondigaugemet}) and the boundary conditions (\ref{bc1})-(\ref{bc2}),  modulo the subgroup of trivial diffeomorphisms.\footnote{ See~\cite{Barnich:2006av} for a related derivation of the BMS algebra in higher dimensions.} 
All such diffeomorphisms are generated by vector fields $\xi$ satisfying the following set of differential equations:
\be \label{bondigauge}
\mathcal{L}_{\xi} g_{rr}=0, \hspace{.5 in } \mathcal{L}_{\xi} g_{ra}=0, \hspace{.5 in } g^{ab}\mathcal{L}_{\xi}g_{ab}=0,  
\ee
\be \label{afbc}
 \mathcal{L}_{\xi}g_{uu}=\bigO(r^{-1}), \hspace{.5 in} \mathcal{L}_{\xi}g_{ur}=\bigO (r^{-2}), \hspace{.5 in} \mathcal{L}_{\xi}g_{ua}=\bigO(1), \hspace{.5 in} \mathcal{L}_{\xi}g_{ab}=\bigO(r).   
\ee
The most general vector field satisfying (\ref{bondigauge})-(\ref{afbc}) takes the form
\be
\xi^u=f(z)+\frac{u}{2m}D_aY^a(z),   
\ee
\be
\xi^a=Y^a(z)-D_b\xi^u \int_r^{\infty}e^{-2\beta}g^{ab}dr',   
\ee
\be
\xi^r=-\frac{r}{2m}[D_a\xi^a -U^aD_a\xi^u].   
\ee
The $Y^a(z)$ are  conformal Killing vectors on the $S^{2m}$, and generate the $SO(2m+1,1)$ transformations of the Poincare group. Transformations with $Y^a=0$ and an arbitrary function $f(z)$ on the sphere are known as supertranslations. 
Near~$\mathcal{I}^+$, they are generated by the vector field
\be \label{vecfield}
\xi=f\p_u -\frac{1}{r}\gamma^{ab}D_a f \p_b +\frac{1}{2m} D^2 f \p_r+\dots   
\ee
In the linearized theory, where we ignore transformations homogenous in metric perturbations (such transformations require quadratic terms in the associated charges), the effect of a supertranslation is to shift $C_{ab}^{(-1)}$ according to  
\be\label{eq:ch1}
\delta C_{ab}^{(-1)}
=\frac{1}{m}D^2f\gamma_{ab} -(D_aD_b+D_bD_a)f,    
\ee
leaving all other $C_{ab}^{(n \geq 0)}$ fixed. Note that  $\delta C_{ab}^{(-1)}=0 $ for the $2m+2$ global translations given by $f(z)\propto 1,\hat{x}^i(z)$.  
A similar analysis holds for the asymptotic symmetry group at $\mathcal{I}^-$. There, supertranslation generators take the asymptotic form
\be
\xi^-=f^-\p_v +\frac{1}{r}\gamma^{ab}D_a f^- \p_b -\frac{1}{2m} D^2 f^- \p_r+\dots   
\ee
and generate the transformation
\be\label{eq:ch2}
\delta D_{ab}^{(-1)}
=-\left[\frac{1}{m}D^2f^-\gamma_{ab} -(D_aD_b+D_bD_a)f^- \right].   
\ee

\subsection{Discussion of the BMS group in higher dimensions}
Previous analyses of higher dimensional asymptotically flat spacetimes have concluded that the BMS group does not exist in higher dimensions and that the appropriate asymptotic symmetry group is the finite dimensional Poincare group \cite{Hollands:2003xp,Hollands:2003ie,Tanabe:2009va,Tanabe:2010rm,Tanabe:2011es,Tanabe:2012fg}. Notable exceptions include \cite{Barnich:2006av,Awada:1985by}, where it was argued that supertranslations do exist in higher dimensions. The source of the apparent discrepancy can be found in the choice of boundary conditions. 
In $2m+2$ spacetime dimensions, the radiative degrees of freedom of the gravitational field enter the metric on the sphere at order $\bigO(r^{2-m})$. As we have seen, supertranslations affect an $\bigO(r)$ change in the metric on the sphere. For $d=4$, these two orders agree, and it is impossible to eliminate the supertranslations without simultaneously eliminating radiative solutions. In higher dimensions, one can consistently set $C^{(-1)}_{ab}=0$ while still allowing for radiative solutions, which have nonzero $C_{ab}^{(m-2)}$. Since the boundary condition  $C^{(-1)}_{ab}=0$ is not supertranslation invariant, it effectively reduces the infinite dimensional BMS group to the finite dimensional Poincare group. 

The definition of an asymptotic symmetry group depends on the boundary conditions of the theory, and the appropriate boundary conditions are often determined by the phenomena under consideration. Therefore, in higher dimensions it is meaningless to discuss the ``correct" asymptotic symmetry group, and one should simply choose the group best adapted to the problem at hand. In four dimensions, the extra supertranslation symmetries were intimately related to the infrared behavior of gravitational scattering amplitudes. Therefore it seems reasonable that if one wishes to study the infrared dynamics of higher dimensional gravity, one should choose the relaxed boundary conditions (\ref{bc1}) which allow for supertranslations. As we will see in sections five and six, this is indeed the case.

\subsection{Boundary conditions and constraints}
In this section we collect a few select formulas which will prove useful for the analysis of Weinberg's soft theorem. The analysis up to this point is completely general and holds in the non-linear theory. In what follows we will focus on the linearized theory for ease of presentation. First, note that linearization effectively eliminates the function $\beta$ in the metric. 
In the nonlinear theory,
\be
R_{rr}= \frac{4m}{r}\p_r \beta+\frac{2m}{r^2}-\frac14 g^{ac}g^{bd}\p_r g_{ab}\p_rg_{cd}.   
\ee
The boundary conditions $R_{rr}=\bigO(r^{-2m-2})$ and  $g_{ur}=\bigO(r^{-2})$ then imply that $\beta$ is quadratic in metric perturbations up to order $\bigO(r^{-2m+1})$. Since these are the only orders of $\beta$ that could appear in the equations (\ref{uuexpand})-(\ref{constraint}) needed for our analysis, $\beta$ may be consistently set to zero along with all other terms quadratic in metric perturbations. Also note that in the linearized theory, the Bondi gauge determinant condition requires that all $C_{ab}^{(n)}$ be traceless.
We then have
\be\label{eq:uu}
R_{uu}=-\frac{1}{r^2}\p_u D^aU_a-\frac{1}{2}r^{-2m}\p_r(r^{2m}\p_rM)  -\frac{m}{r}\p_uM-\frac{1}{2r^2}D^2M,   
\ee
\be\label{eq:ra}
R_{ra}=\frac12r^{-2m}\p_r(r^{2m+2}\p_r(r^{-2}U_a))     +\frac{1}{2r^2}\p_rD^b g_{ba}+r\gamma_{ca}D_bg^{bc},   
\ee
\be\label{eq:ur}
R_{ur}=-\frac{1}{2}r^{-2m}\p_r(r^{2m}\p_r M)-\frac{1}{2r^2}\p_r D^a U_{a}.   
\ee
The corresponding equations in advanced Bondi gauge are:
\be\label{eq:vv}
R_{vv}=-\frac{1}{r^2}\p_v D^aW_a -\frac{1}{2}r^{-2m}\p_r(r^{2m}\p_rM^-)  +\frac{m}{r}\p_vM^-   -\frac{1}{2r^2}D^2M^-  , 
\ee
\be\label{eq:rav}
R_{ra}=-\frac12r^{-2m}\p_r(r^{2m+2}\p_r(r^{-2}W_a))+\frac{1}{2r^2}\p_rD^b g^{-}_{ba}+r\gamma_{ca}D_bg^{-bc}, 
\ee
\be\label{eq:vr}
R_{vr}=\frac{1}{2}r^{-2m}\p_r(r^{2m}\p_r M^-)-\frac{1}{2r^2}\p_r D^a W_{a}.  
\ee
The boundary condition on $R_{uu}$ reads 
\be \label{uuexpand}
  \frac12 [D^2+ n(n+1-2m)] M^{(n)}  + \p_u D^a U_a^{(n)} + m  \p_u M^{(n+1)}  =0,  \hspace{.3 in} 0\leq n \leq 2m-3.   
\ee
The boundary condition on  $R_{ur}$ reads
\be
 -\frac{ n(n+1-2m)}{2}M^{(n)}  +\frac{  (n-1)}{2} D^a U_a^{(n-1)}=0, \hspace{.5 in}  0 \leq n \leq 2m-2.   
\ee
The boundary condition on $R_{ra}$ reads
\be
 \frac{(n+2)(n+1-2m)}{2}U_a^{(n)} - \frac{(n+1)}{2}D^bC_{ba}^{(n-1)}=0, \hspace{.5 in} 0\leq n \leq 2m-2.   
\ee
The null normal to $\mathcal{I}^+$ is given by $n=\p_u-\frac12 \p_r$. The constraint equations take the the form 
\be
n^{\mu}(R_{\mu\nu}-\frac{1}{2}Rg_{\mu \nu}) =8\pi G n^{\mu} T_{\mu\nu}^M.   
\ee 
The boundary conditions (\ref{bcric})-(\ref{bc2}) ensure that $R=\bigO(r^{-2m-1})$. In what follows we never need terms of this order, so we drop the trace term from the constraint equation.   The first nontrivial order of the $u$-constraint equation then reads
\be \label{constraint}
 \frac12 [D^2- 2(m-1)] M^{(2m-2)}  + \p_u D^a U_a^{(2m-2)} + m  \p_u M^{(2m-1)}  + 8\pi G T^{M(2m)}_{uu} =0.   
\ee

\section{The semi-classical scattering problem}
So far, our analysis has treated $\mathcal{I}^+$ and $\mathcal{I}^-$ separately. In order to discuss the symmetries of the gravitational $\mathcal{S}$-matrix, we need to define the semiclassical scattering problem in general relativity and determine how to relate symmetry transformations at $\mathcal{I}^+$ and $\mathcal{I}^-$. In essence, given initial data for the characteristic Cauchy problem at $\mathcal{I}^-$, we must determine the corresponding outgoing data on $\mathcal{I}^+$. One of the most attractive features of the asymptotic analysis based at $\mathcal{I}$ is the ability to solve this problem without making reference to the interior of the spacetime. However, in order to do so, we need to be able to relate data and symmetry transformations on $\mathcal{I}^-$ to the corresponding data and transformations on $\mathcal{I}^+$. Doing so requires us to impose certain regularity conditions at spatial infinity.

\subsection{CK constraint in higher dimensions}

In four dimensions, arbitrary asymptotically flat initial data does not lead to a well-defined scattering problem. In fact, $i^0$ is generically a singular point of the conformal compactification of asymptotically flat spacetimes. This naively precludes the identification of BMS transformations at $\ip$ and $\im$. In four dimensions, the work of Christodoulou and Klainerman (CK) \cite{Christodoulou:1993uv} established necessary bounds on initial data in order to allow for smooth identifications at $i^0$. To our knowledge, no such analysis has been performed in higher dimensions. In four dimensions, the CK conditions played an essential role in connecting $\ip$ to $\im$ and matching gravitational data at $i^0$. We are thus led to impose a ``generalized CK constraint." Specifically, we require that the higher dimensional analog of the magnetic component of the Weyl tensor $C_{\mu \nu \rho \sigma}$ vanishes near the boundaries of $\ip $:

\be
C_{urab}|_{\ip_{\pm}}=\cO(r^{-2}).   
\ee
The $\bigO(r^{-1})$ term in this constraint requires that  
\be
D_a U_b^{(0)} - D_b U_a^{(0)}=0   
\ee
at $\mathcal{I}^+_{\pm}$. The $\bigO(1)$ $R_{ab}$ boundary condition implies  $\p_uC^{(-1)}_{ab}=0$. Combining these two conditions we see that $D^bC^{(-1)}_{ba}=D_ag(z)$ for some function $g(z)$ on the sphere.  The most general solution consistent with Bondi gauge is
\be
C_{ab}^{(-1)} = \frac{1}{m}\gamma_{ab}D^2\psi(z)-2D_a D_b \psi(z),
\ee
with $\psi(z)$ unconstrained. Note that this is simply the requirement that $C_{ab}^{(-1)}$ be pure supertranslation. Under the action of a supertranslation with parameter $f(z)$,  $\psi(z)$ transforms according to $\psi(z)\to \psi(z)+f(z)$.  The function $\psi(z)$ will later be identified as the goldstone mode for spontaneously broken supertranslation symmetry. 
The analogous condition at $\mathcal{I}^-$ yields
\be
D_{ab}^{(-1)}=-\frac{1}{m}\gamma_{ab}D^2\psi^-(z)+2D_a D_b \psi^-(z) . 
\ee

In what follows, we are primarily interested in vacuum to vacuum geometries, and we impose the ``scattering constraints"
\be
M^{(2m-1)}|_{\mathcal{I}^+_+}=M^{-(2m-1)}|_{\mathcal{I}^-_{-}}=0, \hspace{.5 in} C_{ab}^{(2m-3)}|_{\mathcal{I}^+_{\pm}}=D_{ab}^{(2m-3)}|_{\mathcal{I}^-_{\pm}}=0. 
\ee

\subsection{Scattering and matching}
In order to connect $\mathcal{I}^-$ to $\mathcal{I}^+$ we must match data at $i^0$. Following the analysis in \cite{Strominger:2013jfa}, all fields and functions are taken to be continuous along the null generators of $\mathcal{I}$ passing through $i^0$. Due to the antipodal identification of the angular coordinates on $\mathcal{I}^+$ and $\mathcal{I}^-$, the zero modes $C_{ab}^{(-1)}$ and $D_{ab}^{(-1)}$ are matched according to
\be
\psi(z)=\psi^-(z).   
\ee
This identification also allows for a canonical identification of $BMS^+$ and $BMS^-$ transformations according to the rule
\be
f(z)=f^{-}(z),   
\ee
yielding a diagonal BMS subgroup that may be identified as a symmetry of the gravitational $\mathcal{S}$-matrix.

\section{Six dimensional gravity}
In this section we focus on six dimensional asymptotically flat spacetimes. We identify the free radiative data, and collect a number of equations needed for the analysis of Weinberg's soft theorem. 

\subsection{Boundary conditions and constraints}
In six dimensions, the boundary conditions for asymptotically flat spacetimes satisfying the scattering constraints and the generalized CK constraint take the form
\be \begin{split}
g_{uu}=-1+\bigO(r^{-1}), \hspace{.3 in} g_{ur}=-1+\bigO(r^{-2}), \hspace{.3 in} g_{ua}=\bigO(1), \hspace{.3 in} g_{ab}=r^2\gamma_{ab}+\bigO(r),
\\
R_{uu}=\bigO(r^{-4}), \hspace{.2 in} R_{ur}=\bigO(r^{-5}), \hspace{.2 in} R_{ua}=\bigO(r^{-4}), \hspace{.2 in}  R_{rr}=\bigO(r^{-6}), \hspace{.2 in}  R_{ra}=\bigO(r^{-5}),
\\
  R_{ab} = \bigO(r^{-4}), \hspace{.3 in}C_{ab}^{(1)}|_{\mathcal{I}^+_{\pm}}=0,  \hspace{.3 in} C_{urab}|_{\mathcal{I}^+_{\pm}}=\bigO(r^{-2}), \hspace{.3 in} M^{(3)}|_{\mathcal{I}^+_+}=0 . \hspace{.5 in} \end{split}   
\ee 
The $R_{uu}$ boundary conditions take the form
\be
\p_u[D^aU_a^{(0)}+2M^{(1)}]=0, \hspace{.5 in} \frac12[D^2-2]M^{(1)}+\p_u[D^aU_a^{(1)}+2M^{(2)}]=0,   
\ee
while the $R_{ur}$ boundary conditions reduce to
\be
M^{(1)} =0 , \hspace{.5 in }M^{(2)}=-\frac12 D^aU_a^{(1)}.   
\ee
The $R_{ra}$ boundary conditions read 
\be
U_a^{(0)}=-\frac16 D^bC_{ba}^{(-1)} , \hspace{.5 in} U_a^{(1)}=-\frac13 D^bC_{ba}^{(0)}, \hspace{.5 in} U_a^{(2)}=-\frac34 D^bC_{ba}^{(1)}.    
\ee
The $\bigO(r^{-4}) $ $u$-constraint equation reads
\be
\frac12[D^2-2]M^{(2)}+\p_u D^aU_a^{(2)}+2\p_uM^{(3)}=-8\pi G T_{uu}^{M(4)}.  
\ee
$C_{ab}^{(0)}$ is free, unconstrained data. A complete solution of course requires integration of the constraints and equations of motion to all orders.

\subsection{Mode expansions}
The fluctuations of the gravitational field in an asymptotically flat spacetime are determined by the relation $g_{\mu \nu}=\eta_{\mu \nu}+\kappa h_{\mu \nu}$, where $\kappa^2 = 32\pi G$ and $\eta_{\mu \nu}$ is the flat metric. We represent the radiative degrees of freedom of the gravitational field by the mode expansion
\be
h_{\mu \nu}(x)=\sum_{\alpha}\int \frac{d^5q}{(2\pi)^5}\frac{1}{2\omega_q}\left[\e^{*\alpha}_{\mu \nu}a_{\alpha}(\vec{q})e^{iq{ \cdot }x}+\e^{\alpha}_{\mu \nu}a_{\alpha}(\vec{q})^{\dagger}e^{-iq{\cdot }x}				\right].   
\ee
Here $\omega_q=|\vec{q}|$ and  $\e^\a_{\mu\nu}$  are the polarization tensors of the graviton in six dimensions. The commutation relations are given by 
\be
[a_\a(\vec{p}), a_\beta(\vec{q})^\dagger ] = 2\omega_q \delta_{\alpha\beta} (2\pi)^{5} \delta^{5}(\vec{p}-\vec{q}).   
\ee
The free radiative data at $\mathcal{I}^+$ then takes the form 
\be
C_{ab}^{(0)} (u,z) \equiv \kappa\lim_{r\to \infty} \p_a x^\mu \p_b x^\nu h_{\mu\nu} (u+r, r\hat{x}(z)).   
\ee
We evaluate this limit using  the large-$r$ saddle point approximation to obtain the expression
\be \label{modes}
C_{ab}^{(0)}(u,z^a)=-\frac{2\pi^2\kappa}{(2\pi)^5}\p_a \hat{x}^i \p_b \hat{x}^j\sum_{\alpha}\int \omega_q d\omega_q \left[ \e^{*\alpha}_{ij}a_{\alpha}(\omega_q \hat{x})e^{-i\omega_q u}+\e_{ij}^{\alpha}a_{\alpha}(\omega_q \hat{x})^{\dagger}e^{i\omega_q u}				\right].   
\ee
The positive and negative frequency modes are then given by 
\be  \label{modes2}\begin{split}
C_{ab}^{\omega ( 0)}(z)=-\frac{\kappa \omega}{8\pi^2}\p_a \hat{x}^i (z)\p_b \hat{x}^j (z)\sum_{\alpha}\e_{ij}^{*\alpha}a_{\alpha}(\omega \hat{x}(z)),
\\
C_{ab}^{-\omega ( 0)}(z)=-\frac{\kappa \omega}{8\pi^2}\p_a \hat{x}^i (z)\p_b \hat{x}^j (z)\sum_{\alpha}\e_{ij}^{\alpha}a_{\alpha}(\omega \hat{x}(z))^{\dagger}, \end{split}   
\ee
where $\omega >0$ in both formulas. The $\omega \to 0$ limit of these expressions defines a zero mode operator
\be
C_{ab}^{0(0)}=\frac{1}{2}\lim_{\omega \to 0}(	C_{ab}^{\omega (0)} + C_{ab}^{-\omega (0)}		).   
\ee
The asymptotic data at $\mathcal{I}^-$ is given by
\be 
D_{ab}^{(0)}(v,z)=\kappa\lim_{r\to \infty}\p_a x^\mu \p_bx^\nu h_{\mu \nu} (v-r,r\hat{x}^i(z)),   
\ee
which may be decomposed into the positive and negative frequency modes
\be \begin{split}
D_{ab}^{\omega ( 0)}(z)=-\frac{\kappa \omega}{8\pi^2}\p_a \hat{x}^i(z) \p_b \hat{x}^j (z)\sum_{\alpha}\e_{ij}^{*\alpha}a_{\alpha}(-\omega \hat{x}(z)), 
\\
D_{ab}^{-\omega ( 0)}(z)=-\frac{\kappa \omega}{8\pi^2}\p_a \hat{x}^i(z) \p_b \hat{x}^j (z)\sum_{\alpha}\e_{ij}^{\alpha}a_{\alpha}(-\omega \hat{x}(z))^{\dagger}. \end{split}   
\ee
The associated zero mode creation operator is given by
\be
D_{ab}^{0(0)}=\frac12\lim_{\omega \to 0}(D_{ab}^{\omega ( 0)}+D_{ab}^{-\omega ( 0)}).   
\ee

\section{Ward identity from Weinberg's soft theorem}
In this section we use Weinberg's six-dimensional soft graviton theorem to derive a Ward identity for a charge operator constructed from the gravitational field. In the following section, we demonstrate the relationship between this charge operator and the supertranslations described in section 2. 
\subsection{Derivation of Ward identity}
 In six dimensions, Weinberg's soft graviton theorem takes the form
\be \label{weinberg}
\lim_{\omega \to 0} \omega \langle z_{n+1},\dots|a_{\alpha }(q)\mathcal{S}|z_1,\dots \rangle=\frac{\omega \kappa}{2}\left[   \sum_{k=n+1}^{n+n'}\frac{\e^{\alpha}_{\mu \nu}p_k^{\mu}p_k^{\nu}}{p_k\cdot q}-\sum_{k=1}^{n}\frac{\e^{\alpha}_{\mu \nu}p_k^{\mu}p_k^{\nu}}{p_k\cdot q} \right] \langle z_{n+1},\dots|\mathcal{S}|z_1,\dots \rangle.   
\ee 
Here $a_\a(q)$ is a creation operator for an outgoing on-shell graviton with energy $\omega$, polarization $\e^\a_{\mu \nu}$ and momentum $q^{\mu}$. A null momentum vector in six dimensions is determined by an energy $\omega$ and a point $z^a$ on the $S^4$, so we parametrize the soft graviton's momentum as
 \begin{equation} \label{q}
 q^{\mu}=\omega \left[1,\hat{x}^i(z)\right].    \end{equation} 
 Here $\hat{x}^i(z)$ is the embedding of $S^{4}$  into $\mathbb{R}^{5}$ defined previously. The momenta of the massless external particles are similarly given by 
 \begin{equation} \label{p}
 p^\mu_k=E_k\left[1,\hat{x}^i(z_k)\right] .    
 \end{equation}
Thus the in- and out-states are determined by the energy $E_k$ and $\ci^+$ crossing point  $z_k$ for each external particle. For simplicity we suppress internal quantum numbers which are totally decoupled from the analysis. We denote the in- and out-states by
 \be \label{states}
|z_1,\dots,z_n \rangle,\;\;\; \langle z_{n+1},\dots,z_{n+n'}|,   
\ee  
respectively.
Motivated by the form of the radiative modes (\ref{modes2}), we define a function
\be \begin{split}
F^{out}_{ab}(z,z_1,\dots,z_{n+n'})\equiv\omega\p_a \hat{x}^i(z)\p_b \hat{x}^j(z)\sum_{\alpha}\e_{ij}^{*\alpha}\left[   \sum_{k=n+1}^{n+n'}\frac{\e^{\alpha}_{\mu \nu}p_k^{\mu}p_k^{\nu}}{p_k\cdot q}-\sum_{k=1}^{n}\frac{\e^{\alpha}_{\mu \nu}p_k^{\mu}p_k^{\nu}}{p_k\cdot q} \right]
\\=\sum_{k=n+1}^{n+n'}  E_k\p_b P(z,z_k) \p_a\log(1-P(z,z_k))-\sum_{k=1}^{n}   E_k\p_b P(z,z_k) \p_a\log(1-P(z,z_k)).  
\end{split} 
\end{equation}
Here we have used the completeness relation for polarization tensors
\begin{equation}
2\sum_{\alpha}\epsilon_{\alpha}^{* ij}(\vec{q})\epsilon^{kl}_{\alpha}(\vec{q})=\pi^{ik}\pi^{jl}+\pi^{il}\pi^{jk}-\frac12 \pi^{ij}\pi^{kl},\;\;\;\pi^{ij}=\delta^{ij}-\frac{q^iq^j}{\vec{q}^2},   
\end{equation}
energy and momentum conservation 
\be
\sum_{k=n+1}^{n+n'}  E_k-\sum_{k=1}^{n} E_k=0, \hspace{.5 in} \sum_{k=n+1}^{n+n'}  E_k\hat{x}^i(z_k)-\sum_{k=1}^{n} E_k\hat{x}^i(z_k)=0,  
\ee
and defined a function\footnote{$P$ is known as the invariant distance on the $S^4$, and is related to the cosine of the geodesic distance.}
\be
P(z,z_k)\equiv  \hat{x}_i (z)\hat{x}^i (z_k).  
\ee 
We then use $F^{out}_{ab}(z,z_1,\dots,z_{n+n'})$ (abbreviated $F^{out}_{ab}(z;z_k)$) to relate Weinberg's soft theorem to the zero mode insertion:
\be\label{0insert1}
\langle z_{n+1},\dots |C_{ab}^{0(0)}(z)\mathcal{S}|z_1,\dots \rangle =-\frac{\kappa^2}{2(4\pi)^2}F_{ab}^{out}(z;z_k)\langle z_{n+1},\dots| \mathcal{S}|z_1,\dots \rangle.  
\ee
Note that $F_{ab}^{out}(z;z_k)$ obeys the differential equation
\be \label{diffeq}
\sqrt{\gamma}[D^2-2]D^aD^bF_{ab}^{out}=3(4\pi)^2\left[	\sum_{k=n+1}^{n+n'}E_k \delta^4(z-z_k)-\sum_{k=1}^{n}E_k \delta^4(z-z_k)	\right].  
\ee
We can similarly consider Weinberg's soft theorem for an incoming soft graviton
\be \notag
\lim_{\omega \to 0} \omega \langle z_{n+1},\dots|\mathcal{S}a_{\alpha }(q)^{\dagger}|z_1,\dots \rangle=-\frac{\omega\kappa}{2}\left[   \sum_{k=n+1}^{n+n'}\frac{\epsilon^{*\alpha}_{\mu \nu}p_k^{\mu}p_k^{\nu}}{p_k\cdot q}-\sum_{k=1}^{n}\frac{\epsilon^{*\alpha}_{\mu \nu}p_k^{\mu}p_k^{\nu}}{p_k\cdot q} \right]  \langle z_{n+1},\dots|\mathcal{S}|z_1,\dots \rangle.  
\ee 
This can similarly be rewritten
\be \label{0insert2}
\langle z_{n+1},\dots|\mathcal{S}D_{ab}^{0(0)}(z)|z_1,\dots \rangle=\frac{\kappa^2}{2(4\pi)^2}F_{ab}^{in}(z;z_k)\langle z_{n+1},\dots | \mathcal{S}|z_1,\dots \rangle,  
\ee
where
\begin{align}
&F_{ab}^{in}(z,z_1,\dots,z_{n+n'}) \notag
=\\&-\sum_{k=n+1}^{n+n'}  E_k\p_b P(z,z_k) \p_a\log(1+P(z,z_k))+\sum_{k=1}^{n}   E_k\p_b P(z,z_k) \p_a\log(1+P(z,z_k)) .  
\end{align}
Combining equations (\ref{0insert1}),  (\ref{diffeq}) and  (\ref{0insert2}) we obtain the identity
\begin{align} \begin{split}
-\frac{1}{3\kappa^2}\int d^4z \sqrt{\gamma}f(z)(D^2-2)D^aD^b\langle z_{n+1},\dots |C_{ab}^{0(0)}(z)\mathcal{S}|z_1,\dots \rangle \\ 
+\frac{1}{3\kappa^2}\int d^4z \sqrt{\gamma}f^-(z)(D^2-2)D^aD^b\langle z_{n+1},\dots |\mathcal{S}D_{ab}^{0(0)}(z)|z_1,\dots \rangle\\
=\left[ \sum_{k=n+1}^{n+n'}E_k f(z_k)-\sum_{k=1}^{n}E_k f(z_k)\right] \langle z_{n+1},\dots |\mathcal{S}|z_1,\dots \rangle	 \end{split} 
\end{align}
for an arbitrary function $f(z)$ on the sphere. This relation can be rewritten as a Ward identity
\be
\langle z_{n+1},\dots |Q^+\mathcal{S}-\mathcal{S}Q^-|z_1,\dots \rangle=0,  
\ee
where $Q^{\pm}$ have been decomposed into hard and soft parts
\be\label{charges}
Q^{\pm}=Q_H^{\pm} +Q_S^{\pm}.  
\ee
The action of the hard charges is defined so that 
\be \label{qhard}
Q_H^{-}|z_1,\dots \rangle=\sum_{k=1}^{n} E_k f(z_k)|z_1,\dots \rangle, \;\;\;\;\; \langle z_{n+1},\dots | Q_{H}^+ =\langle z_{n+1},\dots | \sum_{k=n+1}^{n+n'}E_k f(z_k),  
\ee
while the soft charges take the form 
\begin{align} \begin{split}
Q_S^+=\frac{1}{3\kappa^2}\int d^4z \sqrt{\gamma}f(z)(D^2-2)D^aD^bC_{ab}^{0(0)}(z), \\
Q_S^-=\frac{1}{3\kappa^2}\int d^4z \sqrt{\gamma}f^-(z)(D^2-2)D^aD^bD_{ab}^{0(0)}(z). \end{split} 
\end{align}

\section{From Ward identity to BMS supertranslations}
The charges (\ref{charges}) commute with the $\mathcal{S}$-matrix and represent a symmetry of the theory. In this section we argue  that, given the zero-mode bracket postulated below,   the symmetry generated by these charges is none other than the supertranslation symmetry encountered in section 2.

\subsection{Action of the matter charges }
The form of the hard charges may be deduced from (\ref{qhard}), yielding  
\be
Q_{H}^+=\lim_{r\to \infty} r^4 \int_{\mathcal{I}^+}du d^4z\sqrt{\gamma}\; f(z) T_{uu}^M(u,r,z),  
\ee
\be
Q_{H}^-=\lim_{r\to \infty}r^4\int_{\mathcal{I}^-}dv d^4z \sqrt{\gamma} f^-(z)T_{vv}^M(v,r,z).  
\ee
These expressions can be rewritten in the form 
\be\label{cgen}
Q_{H}^+=\lim_{\Sigma\to \ip}  \int_{\Sigma} d\Sigma\;  \xi^\mu n^\nu_{\Sigma}T_{\mu\nu}^M,\;\;\;\;\;\;\;\;\;\;\; Q_{H}^-=\lim_{\Sigma\to \im}  \int_{\Sigma} d\Sigma\;  \xi^\mu n^\nu_{\Sigma}T_{\mu\nu}^M.  
\ee
Here $\Sigma$ is a space-like Cauchy surface, $n_\Sigma$ is a unit normal to $\Sigma$, and $\xi$ is the BMS vector field (\ref{vecfield}). Written in this form, it is clear that the hard charges generate supertranslations on the asymptotic states. Standard commutation relations for the matter fields confirm that the quantities (\ref{cgen}) generate the large diffeomorphisms for any matter field coupled to gravity. 

Note that because we consider the linearized theory, the gravitational field does not appear in the hard charges. In the full nonlinear theory, when we allow for external graviton states with non-zero momentum, we expect new contributions to the hard charge quadratic in the gravitational field. These additional terms characterize the energy and momentum flux  carried by gravitational radiation through null infinity, and serve to generate transformations of the metric that are proportional to metric perturbations (transformations which we neglect in the linearized theory).

\subsection{Action of the gravitational charges}

Supertranslations are by definition large diffeomorphisms, and general relativity is diffeomorphism invariant if and only if all fields transform under the diffeomorphisms. Therefore, it is intuitively clear that the charges (\ref{charges}) must generate supertranslations of the gravitational field. We can make this relationship precise by using the boundary conditions and constraints of section 4.1 to rewrite the charges as boundary integrals
\be\label{bcharges}
\begin{split} 
Q^+=\frac{1}{4\pi G}\int_{\mathcal{I}^+_-}d^4z\sqrt{\gamma} f(z)M^{(3)}(z),\\
Q^-=\frac{1}{4\pi G}\int_{\mathcal{I}^-_+}d^4z\sqrt{\gamma} f^-(z)M^{-(3)}(z).  
\end{split}
\ee

These expressions resemble the supertranslation generators encountered in the four dimensional case \cite{He:2014laa}. In order to claim that they generate supertranslations of the gravitational field, we need to discuss the symplectic form for the gravitational free data.

\subsection{Brackets for the free data}
The commutator for the radiative degrees of freedom of the gravitational field is familiar from four dimensions and can be derived from the plane wave mode expansion. It is given by 
\be
[C^{(0)}_{ab}(u,z),\p_{u'}C_{cd}^{(0)}(u',z')]=i\frac{\kappa^2}{4} \frac{\delta(u-u')\delta^4(z-z')}{\sqrt{\gamma}}\left[\gamma_{ac} \gamma_{bd} +\gamma_{ad} \gamma_{bc} -\frac12 \gamma_{ab} \gamma_{cd}\right].  
\ee
This expression does not determine the zero-mode brackets. We postulate the simple form
\be
[M^{(3)}(z),\psi(z')]=4\pi i G  \frac{\delta^4(z-z')}{\sqrt{\gamma}}.  
\ee
 This reproduces (\ref{eq:ch1}) and defines a symplectic form on the extended gravitational phase space. It closely resembles the analogous zero-mode bracket in QED \cite{Kapec:2014zla}.

\section{Generalization to arbitrary even-dimensional spacetime}

The results of the preceding sections generalize to arbitrary even-dimensional asymptotically flat spacetimes. In this section, we outline the derivation of the supertranslation Ward identity for $d=2m+2$ dimensional spacetime.
 The perturbations of the asymptotically flat gravitational field are defined by $g_{\mu \nu}=\eta_{\mu \nu}+\kappa h_{\mu \nu}$ as in six dimensions. The plane wave expansion takes the form
\begin{equation}\label{2mmode}
h_{\mu\nu}(x)=\sum_{\alpha}\int \frac{d^{2m+1}q}{(2\pi)^{2m+1}}\frac{1}{2\omega_q}\left[	\epsilon_{\mu\nu}^{*\alpha}(\vec{q})a_{\alpha}(\vec{q})e^{iq\cdot x} + \epsilon_{\mu\nu}^{\alpha}(\vec{q})a_{\alpha}(\vec{q})^{\dagger}e^{-iq\cdot x}		\right].  
\end{equation}
Here $\omega_q=|\vec{q}\hspace{.02 in} |$ and  $\a$ labels the polarizations of the graviton.
The operator $a_\a (\vec{q})^\dagger$ is a graviton  creation operator satisfying the commutation relations 
\be
[a_\a(\vec{p}), a_\beta(\vec{q})^\dagger ] = 2\omega_q \delta_{\alpha\beta} (2\pi)^{2m+1} \delta^{2m+1}(\vec{p}-\vec{q}).  
\ee
The leading term in the large-$r$ expansion of (\ref{2mmode})  yields an expression for the radiative degrees of freedom of the gravitational field near $\mathcal{I}^+$. The positive and negative frequency modes take the form
\begin{equation}
C^{\omega(m-2)}_{ab}(z)=\frac{ (-i)^m\omega^{m-1}  \kappa }{2(2\pi)^{m}}\p_a \hat{x}^{j}(z)\p_b \hat{x}^{k}(z)\sum_{\alpha} 	 \epsilon_{jk}^{*\alpha}a_{\alpha}(\omega \hat{x}(z)),   
\end{equation}
\begin{equation}
C^{-\omega(m-2)}_{ab}(z)=\frac{ i^m\omega^{m-1}  \kappa }{2(2\pi)^{m}}\p_a \hat{x}^{j}(z)\p_b \hat{x}^{k}(z)\sum_{\alpha}	 \epsilon_{jk}^{\alpha}a_{\alpha}(\omega\hat{x}(z))^{\dagger}.   
\end{equation}
The corresponding zero mode operator is defined to be
\begin{equation}
C_{ab}^{0(m-2)}=\frac{1}{2}\lim_{\omega \to 0} (i\omega)^{2-m}\left[	C^{\omega(m-2)}_{ab}+(-1)^mC^{-\omega(m-2)}_{ab}		\right].   
\end{equation}
In terms of this zero mode operator, Weinberg's soft theorem (\ref{weinberg}) takes the form

\begin{equation} \label{out2}
\langle z_{n+1},\dots|  C^{0(m-2)}_{ab}(\omega,z) \mathcal{S} | z_1,\dots\rangle = -\frac{(-1)^m\kappa^2}{8(2\pi)^m}F^{out}_{ab}(z;z_k) \langle z_{n+1},\dots| \mathcal{S} | z_1,\dots\rangle .   
\end{equation}
The soft factor 
\begin{equation} \begin{split}
F^{out}_{ab}(z,z_1,\dots,z_{n+n'})\equiv \partial_a\hat{x}^{i}  \partial_b\hat{x}^{j}  \omega  \sum_{\alpha}  \epsilon^{*\alpha}_{ij} \left[     \sum_{k=n+1}^{n+n'} \frac{ \epsilon^{\alpha}_{\mu\nu}p^\mu_k p^\nu_k}{p_k \cdot q}-  \sum_{k=1}^{n}\frac{ \epsilon^{\alpha}_{\mu\nu}p^\mu_k p^\nu_k }{p_k \cdot q}   \right]   
\\  
=\sum_{k=n+1}^{n+n'}  E_k \p_b P(z,z_k)\p_a\log(1-P(z,z_k))-\sum_{k=1}^{n}  E_k \p_b P(z,z_k)\p_a\log(1-P(z,z_k)) \end{split}
\end{equation}
satisfies the dimension-dependent differential equation
\begin{equation}\label{softgen}
\begin{split}
(-1)^{m}\sqrt{\gamma}\prod_{l=m+1}^{2m-1}[D^2-(2m-l)(l-1)]D^aD^b F^{out}_{ab}  \\
=(2m-1)\Gamma(m)2^{m}(2\pi)^m\left[  \sum_{k=n+1}^{n+n'} E_k\delta^{2m}(z-z_k)  -  \sum_{k=1}^n E_k\delta^{2m}(z-z_k)\right].  
\end{split}
\end{equation}
{Here we have used the completeness relation for polarization tensors
\begin{equation}
2\sum_{\alpha}\epsilon_{\alpha}^{* ij}(\vec{q})\epsilon^{kl}_{\alpha}(\vec{q})=\pi^{ik}\pi^{jl}+\pi^{il}\pi^{jk}-\frac{1}{m} \pi^{ij}\pi^{kl},\;\;\;\pi^{ij}=\delta^{ij}-\frac{q^iq^j}{\vec{q}^2},   
\end{equation}
along with energy and momentum conservation.
}
The in-modes take the form
{\begin{equation}
D^{\omega(m-2)}_{ab}(z)=\frac{ i^m\omega^{m-1}  \kappa }{2(2\pi)^{m}}\p_a \hat{x}^{j}(z)\p_b \hat{x}^{k}(z)\sum_{\alpha} 	 \epsilon_{jk}^{*\alpha}a_{\alpha}(-\omega \hat{x}(z)),   
\end{equation}
\begin{equation}
D^{-\omega(m-2)}_{ab}(z)=\frac{ (-i)^m\omega^{m-1}  \kappa }{2(2\pi)^{m}}\p_a \hat{x}^{j}(z)\p_b \hat{x}^{k}(z)\sum_{\alpha}	 \epsilon_{jk}^{\alpha}a_{\alpha}(-\omega\hat{x}(z))^{\dagger},  
\end{equation}
}
and the associated zero mode operator is
\begin{equation}
D_{ab}^{0(m-2)}=\frac{1}{2}\lim_{\omega \to 0} (i\omega)^{2-m}\left[	D^{\omega(m-2)}_{ab}+(-1)^mD^{-\omega(m-2)}_{ab}		\right]. 
\end{equation}
The soft theorem for an incoming soft graviton may be rewritten as 
\begin{equation} \label{outmodes}
\langle z_{n+1},\dots|  \mathcal{S}D^{0(m-2)}_{ab}(z) | z_1,\dots\rangle = \frac{\kappa^2}{8(2\pi)^m}F^{in}_{ab}(z;z_k) \langle z_{n+1},\dots| \mathcal{S} | z_1,\dots\rangle , 
\end{equation}
where 
\begin{align}
&F^{in}_{ab}(\w,z_1,\dots,z_{n+n'}) \notag
=\\&   -\sum_{k=n+1}^{n+n'}  E_k \p_b P(z,z_k)\p_a\log(1+P(\w,z_k))+\sum_{k=1}^{n}  E_k\p_b P(z,z_k) \p_a\log(1+P(\w,z_k)) .
\end{align}

After applying  (\ref{softgen}) to equations (\ref{out2}) and (\ref{outmodes}), we may integrate against an arbitrary function $f(z)$ on the sphere to obtain the Ward identity
\be\label{wardid2m}
\langle z_{n+1}\dots |\left(Q^+ \mathcal{S} -\mathcal{S}   Q^-\right)|z_1, \dots \rangle=0.  
\ee
The charges $Q^\pm= Q_H^\pm + Q_S^{\pm}$ commute with the $\mathcal{S}$-matrix and induce infinitesimal symmetry transformations on   $\mathcal{I}^{\pm}$ states. $Q_H^{\pm}$ is defined by its action on the asymptotic states: 
\be \label{qhard2}
Q^-_H|z_1,\dots \rangle = \sum_{k=1}^n E_k\;f(z_k)  |z_1,\dots\rangle,\;\;\  \langle z_{n+1},\dots |Q_{H}^+ =   \langle z_{n+1},\dots| \sum_{k=n+1}^{n+n'} E_k\; f(z_k).   
\ee 
The soft charges are given by
\be\label{scgen}
Q^+_S =  \frac1{(2m-1)\kappa^2} \frac{2^{2-m}} {\Gamma(m)}    \int  d^{2m}z \sqrt{\gamma} \;f(z)  \prod\limits^{2m-1}_{l=m+1}  (D^2  -(2m-l)(l-1) ) D^a D^b  C^{0(m-2)}_{ab} ,
\ee
\be\label{scgen}
Q^-_S =  \frac{(-1)^m}{(2m-1)\kappa^2} \frac{2^{2-m}} {\Gamma(m)}     \int  d^{2m}z \sqrt{\gamma} \; f^-(z)  \prod\limits^{2m-1}_{l=m+1}  (D^2  -(2m-l)(l-1) ) D^a  D^b D^{0(m-2)}_{ab} . 
\ee
The hard charges $Q_{H}^{\pm}$ can be written in terms of the matter stress-energy tensor
\be \notag Q^+_H=\lim\limits_{r\to \infty} r^{2m}  \int_{\ip}dud^{2m}z\sqrt{\gamma}f(z) T^M_{uu}(u,r,z),  \;\;\;\;Q^-_H=\lim\limits_{r\to \infty} r^{2m}  \int_{\im}dvd^{2m}z\sqrt{\gamma}f^-(z) T^M_{vv}(v,r,z).
\ee
This operator induces an infinitesimal supertranslation with parameter $f(z)$ when acting on the matter fields. 
The boundary conditions and constraints of section 2.5, combined with the generalized CK constraint and the scattering constraints, allow us to write the total charge $Q^\pm = Q_H^{\pm}+Q_S^{\pm}$ as a boundary integral
\be\label{charge_in_d} \begin{split} 
Q^+ =\frac{4m}{\kappa^2} \lim_{r\to \infty} r^{2m-1} \int_{\ip_-} d^{2m}z \;\sqrt{\gamma} \; f(z)\; M(z),   \\
Q^- =\frac{4m}{\kappa^2} \lim_{r\to \infty} r^{2m-1} \int_{\im_+} d^{2m}z \;\sqrt{\gamma} \; f^-(z)\; M^-(z).
\end{split}
\ee
A straightforward generalization of the brackets of section 6.3 can then be used to demonstrate that (\ref{charge_in_d}) generates supertranslations on the matter and gravitational fields.

\section{Conclusions and open questions}

In this paper we considered the asymptotic symmetry group of even-dimensional asymptotically flat spacetimes. Using less restrictive boundary conditions than those previously considered in the literature, we demonstrated that the BMS group naturally arises as the asymptotic symmetry group of asymptotically flat spacetimes in any even dimension. Motivated by the recently discovered correspondence between soft theorems and asymptotic symmetry groups, we considered Weinberg's soft graviton theorem in even-dimensional spacetime and used it to derive a Ward identity for a set of symmetry transformations. Reasonable, physically motivated boundary conditions and a natural extension of the symplectic form allowed us to identify these symmetry transformations as supertranslations. This result further solidifies the general correspondence between soft theorems and asymptotic symmetry groups. 

It would be worthwhile to tackle the hard metric fluctuations at quadratic order and thereby extend the analysis to the full nonlinear theory. It would also be interesting to consider the odd-dimensional case, where special properties of radiating solutions make the conformal methods usually employed in four dimensions essentially useless. Our analysis is carried out at tree level, and while we expect that the leading soft factor is not renormalized (as in four dimensions), it would be useful to explicitly verify this. One could also extend the analysis to allow for massive external states.

\section*{Acknowledgements}
We are  grateful  to T. Adamo, T. He, D. Jafferis, P. Mitra, H. Ooguri,  A. Porfyriadis, M. Schwartz and A. Zhiboedov for useful conversations. This work was supported in part by NSF grant 1205550 and the Fundamental Laws Initiative at Harvard.
The work of V.L. is supported in part by DOE grant DE-SC0011632 and  the Sherman Fairchild scholarship.  The work of S.P. is supported in part by the Smith Fellowship.

\end{document}